\documentclass[epj]{webofc}
\usepackage[utf8]{inputenc}
\usepackage[varg]{txfonts}   
\usepackage{booktabs}
\usepackage{xcolor}
\definecolor{darkred}{rgb}{0.4,0.0,0.0}
\definecolor{darkgreen}{rgb}{0.0,0.4,0.0}
\definecolor{darkblue}{rgb}{0.0,0.0,0.4}
\usepackage[bookmarks,linktocpage,colorlinks,
    linkcolor = darkred,
    urlcolor  = darkblue,
    citecolor = darkgreen]{hyperref}
\wocname{EPJ Web of Conferences}
\woctitle{Lattice2017}
\DeclareMathOperator{\tr}{tr}

\begin{document}
%
\selectlanguage{english}
\title{%
One-loop perturbative coupling of $A$ and~$A_\star$
through the chiral overlap operator
}
\author{%
\firstname{Hiroki} \lastname{Makino}\inst{1} \and
\firstname{Okuto} \lastname{Morikawa}\inst{1}\fnsep\thanks{Speaker, %
\email{o-morikawa@phys.kyushu-u.ac.jp}} \and
\firstname{Hiroshi}  \lastname{Suzuki}\inst{1}\fnsep\thanks{%
 Acknowledges partial support by JSPS Grants-in-Aid for Scientific
 Research Grant Number~JP16H03982.}
}
\institute{%
Department of Physics, Kyushu University
744 Motooka, Nishi-ku, Fukuoka, 819-0395, Japan
}
\abstract{%
Recently, Grabowska and Kaplan constructed a four-dimensional lattice
formulation of chiral gauge theories on the basis of the chiral overlap
operator. At least in the tree-level approximation, the left-handed fermion
is coupled only to the original gauge field~$A$, while the right-handed one
is coupled only to the gauge field~$A_\star$, a deformation of~$A$ by the
gradient flow with infinite flow time. In this paper, we study the fermion
one-loop effective action in their formulation. We show that the continuum
limit of this effective action contains local interaction terms between $A$
and~$A_\star$, even if the anomaly cancellation condition is met. These
non-vanishing terms would lead an undesired perturbative spectrum in the
formulation.
}
\maketitle
\section{Introduction and discussion}
\label{sec:1}

Recently, Grabowska and Kaplan proposed a four-dimensional lattice
formulation of chiral gauge theories~\cite{Grabowska:2016bis}. This
formulation is based on the so-called overlap operator, which can be obtained
from their five-dimensional domain-wall
formulation~\cite{Grabowska:2015qpk}\footnote{As a closely related
six-dimensional domain-wall formulation, see~Ref.~\cite{Fukaya:2016ofi}.} by
the traditional way~\cite{Neuberger:1997bg,Vranas:1997da,Kikukawa:1999sy}. In
this formulation, along the fifth dimension, the original gauge field $A$ is
deformed by the gradient flow~\cite{Narayanan:2006rf,Luscher:2009eq,%
Luscher:2010iy,Luscher:2011bx} for infinite flow time. Since the gradient
flow preserves the gauge covariance, this formulation is manifestly gauge
invariant, \textit{even if the anomaly cancellation condition is not met}.
Although there is a subtlety associated with the topological charge~\cite{%
Grabowska:2015qpk,Grabowska:2016bis,Okumura:2016dsr,Makino:2016auf,%
Hamada:2017tny}, the smeared gauge field after the infinite-flow time,
$A_\star$, only to which the right-handed (invisible) fermion would be
coupled, can be basically considered as pure gauge
(see~Appendix~\ref{sec:A}). Then one would regard their setup as the system
of the left-handed fermion interacting with the gauge field~$A$;\footnote{%
Grabowska and Kaplan's formulation is a modification of that of
\'Alvarez-Gaum\'e and Ginsparg~\cite{AlvarezGaume:1983cs}. The latter
takes~$A_\star = 0$ identically without the gradient flow and it breaks the
gauge invariance.} this picture was however confirmed only in the tree-level
approximation~\cite{Grabowska:2016bis}. It is thus a crucial problem whether
radiative corrections induce the physical coupling of the right-handed
fermion or not.

First, let us see the tree-level decoupling between the physical and
invisible sectors. So far, only when the transition of the flowed gauge field
along the fifth dimension is abrupt, the four-dimensional lattice Dirac
operator has been obtained as an explicit form; this is referred to as the
chiral overlap operator~$\Hat{\mathcal{D}}_\chi$. The
operator~$\Hat{\mathcal{D}}_\chi$ is given by~\cite{Grabowska:2016bis}
\begin{equation}
   a\Hat{\mathcal{D}}_\chi
   =1+\gamma_5\left[1-(1-\epsilon_\star)\frac{1}{\epsilon\epsilon_\star+1}
   (1-\epsilon)\right],
\label{eq:(1.1)}
\end{equation}
where~$a$ is the lattice spacing, and~$\epsilon$ ($\epsilon_\star$) is
the sign function~\cite{Neuberger:1997fp, Neuberger:1998wv}
\begin{equation}
 \epsilon\equiv\frac{H_w(A)}{\sqrt{H_w(A)^2}}\qquad
 \left(\epsilon_\star\equiv\frac{H_w(A_\star)}{\sqrt{H_w(A_\star)^2}}\right),
\label{eq:(1.2)}
\end{equation}
of the Hermitian Wilson Dirac operator
\begin{equation}
 H_w=\gamma_5\left[\frac{1}{2}\gamma_\mu(\nabla_\mu+\nabla_\mu^*)
	      -\frac{1}{2}a\nabla_\mu\nabla_\mu^*-m\right],
 \label{eq:(1.3)}
\end{equation}
where~$m$ is the parameter of the domain-wall height, and~$\gamma_\mu$ is the
Dirac matrix. In this expression, $\nabla_\mu$ is the forward gauge covariant
lattice derivative and~$\nabla_\mu^*$ is the backward one. With the
assumption of abruptness, this Dirac operator depends on the two gauge
fields, $A$ and~$A_\star$. In the classical continuum
limit~\cite{Grabowska:2016bis},
\begin{equation}
   am\Hat{\mathcal{D}}_\chi\stackrel{a\to0}{\to}
   \gamma_\mu D_\mu(A)P_-+\gamma_\mu D_\mu(A_\star)P_+,
\label{eq:(1.4)}
\end{equation}
where~$D_\mu(A)$ ($D_\mu(A_\star)$) is the covariant derivative defined
with respect to~$A$ ($A_\star$), and~$P_\pm = (1\pm\gamma_5) / 2$ are the
chirality projection operators. Therefore, the coupling between the gauge
fields, $A$ and~$A_\star$, is not produced in the tree-level approximation.

Let us study how the decoupling between $A$ and~$A_\star$ is modified under
radiative corrections. The fermion one-loop effective action is defined by
\begin{equation}
   \ln\mathcal{Z}[A,A_\star]
   \equiv\ln\int\prod_x\left[d\psi(x)d\Bar{\psi}(x)\right]
   \exp\left[-a^4\sum_x\Bar{\psi}(x)\Hat{\mathcal{D}}_\chi\psi(x)\right],
\label{eq:(1.5)}
\end{equation}
where~$A$ and~$A_\star$ are regarded as independent non-dynamical variables.
To investigate the (de)coupling, two infinitesimal variations~$\delta$
and~$\delta_\star$ are introduced such that $\delta$ acts only on~$A$ but not
on~$A_\star$,
\begin{equation}
   \delta A\neq0,\qquad\delta A_\star\equiv0,
\label{eq:(1.6)}
\end{equation}
and~$\delta_\star$ acts in an opposite way,
\begin{equation}
   \delta_\star A\equiv0,\qquad\delta_\star A_\star\neq0.
\label{eq:(1.7)}
\end{equation}
Then, we will find that in the continuum limit a double variation of the
effective action is given as
\begin{equation}
   \delta\delta_\star\ln\mathcal{Z}[A,A_\star]
   =-\int d^4x\,\mathcal{L}(A,A_\star;\delta A,\delta_\star A_\star),
\label{eq:(1.8)}
\end{equation}
where~$\mathcal{L}(A,A_\star;\delta A,\delta_\star A_\star)$ is a
\textit{local} polynomial of its arguments and their spacetime derivatives.

To find a possible implication of~Eq.~\eqref{eq:(1.8)}, we take
\textit{gauge variations} as~$\delta$ and~$\delta_\star$:
\begin{align}
   \delta^\omega A_\mu(x)&\equiv\partial_\mu\omega(x)+[A_\mu(x),\omega(x)],&
   \delta^\omega A_{\star\mu}(x)&=0,
\label{eq:(1.9)}
\\
   \delta_\star^\omega A_{\star\mu}(x)
   &\equiv\partial_\mu\omega(x)+[A_{\star\mu}(x),\omega(x)],&
   \delta_\star^\omega A_\mu(x)&=0.
\label{eq:(1.10)}
\end{align}
Since, as a property of the gradient flow, the two gauge fields~$A$
and~$A_\star$ transform in the same way under the gauge transformation, the
gauge invariance of the effective action implies
\begin{equation}
   (\delta^\omega+\delta_\star^\omega)\ln\mathcal{Z}[A,A_\star]=0
   \Rightarrow
   \delta(\delta^\omega+\delta_\star^\omega)\ln\mathcal{Z}[A,A_\star]=0.
\label{eq:(1.11)}
\end{equation}
Therefore, using~Eq.~\eqref{eq:(1.8)}, we can obtain
\begin{equation}
 \delta\delta^\omega\mathcal{Z}[A,A_\star]
  =-\delta\delta_\star^\omega\mathcal{Z}[A,A_\star]
  =\int d^4x\,\mathcal{L}(A,A_\star;\delta A,\delta_\star^\omega A_\star).
\label{eq:(1.12)}
\end{equation}

Now, let us assume that $A_\star$ becomes pure gauge under the gradient flow
with infinite flow time (see~Appendix~\ref{sec:A}):
\begin{equation}
 A_\star = g^{-1} dg.
\label{eq:(1.13)}
\end{equation}
Then the gauge transformation~$A^{g^{-1}}$ makes~$A_\star = 0$, where
\begin{equation}
 A^g = g^{-1} (d + A) g.
\label{eq:(1.14)}
\end{equation}
That is, we can impose the~$A_\star = 0$ gauge on~Eq.~\eqref{eq:(1.12)}
\begin{equation}
   \delta\delta^\omega\ln\mathcal{Z}[A,0]
   =\int d^4x\,
   \mathcal{L}(A,A_\star=0;\delta A,
   \delta_\star^\omega A_\star|_{A_\star=0}).
\label{eq:(1.15)}
\end{equation}
We will see below that the right-hand side does not vanish even if the
anomaly cancellation condition is met.

It will be shown in the next section that~$\ln\mathcal{Z}[A,0]$ has the term
\begin{equation}
   \ln\mathcal{Z}[A,0]
   = \int d^4x\, \frac{f_0}{2a^2} \tr A_\mu A_\mu
      + \cdots ,
\label{eq:(1.16)}
\end{equation}
thus the mass term~$\tr A_\mu A_\mu$ is produced in the one-loop level. The
propagator of the gauge potential in this~$A_\star=0$ thus has the structure,
\begin{equation}
 \left\langle A_\mu^a(x) A_\nu^a(y) \right\rangle
  = g_0^2\, \delta^{ab} \int \frac{d^4p}{(2\pi)^4}\, e^{i p (x - y)}
  \left(\delta_{\mu\nu} - \frac{p_\mu p_\nu}{p^2}\right)
  \frac{1}{p^2 + m_A^2 + \cdots} ,
\label{eq:(1.17)}
\end{equation}
where we have defined the mass parameter~$m_A$ as
\begin{equation}
 m_A^2 = g_0^2 \frac{f_0}{2a^2} .
\label{eq:(1.18)}
\end{equation}
Therefore, the perturbative spectrum is modified in a \textit{weird} way;
this would not be what we want to obtain for chiral gauge theories. Since
these effects in the one-loop effective action~\eqref{eq:(1.16)} should be
removed by local counterterms, the formulation of Grabowska and Kaplan will
be undesirable as a non-perturbative formulation of chiral gauge theories.
Then their formulation with the abrupt transition should be improved in some
possible way.

\section{Explicit forms of $\mathcal{L}$ and $\delta^\omega\ln\mathcal{Z}$}
\label{sec:2}

\begin{figure}[ht]
 \begin{minipage}{0.45\columnwidth}
  \centering
  \includegraphics[width=\columnwidth]{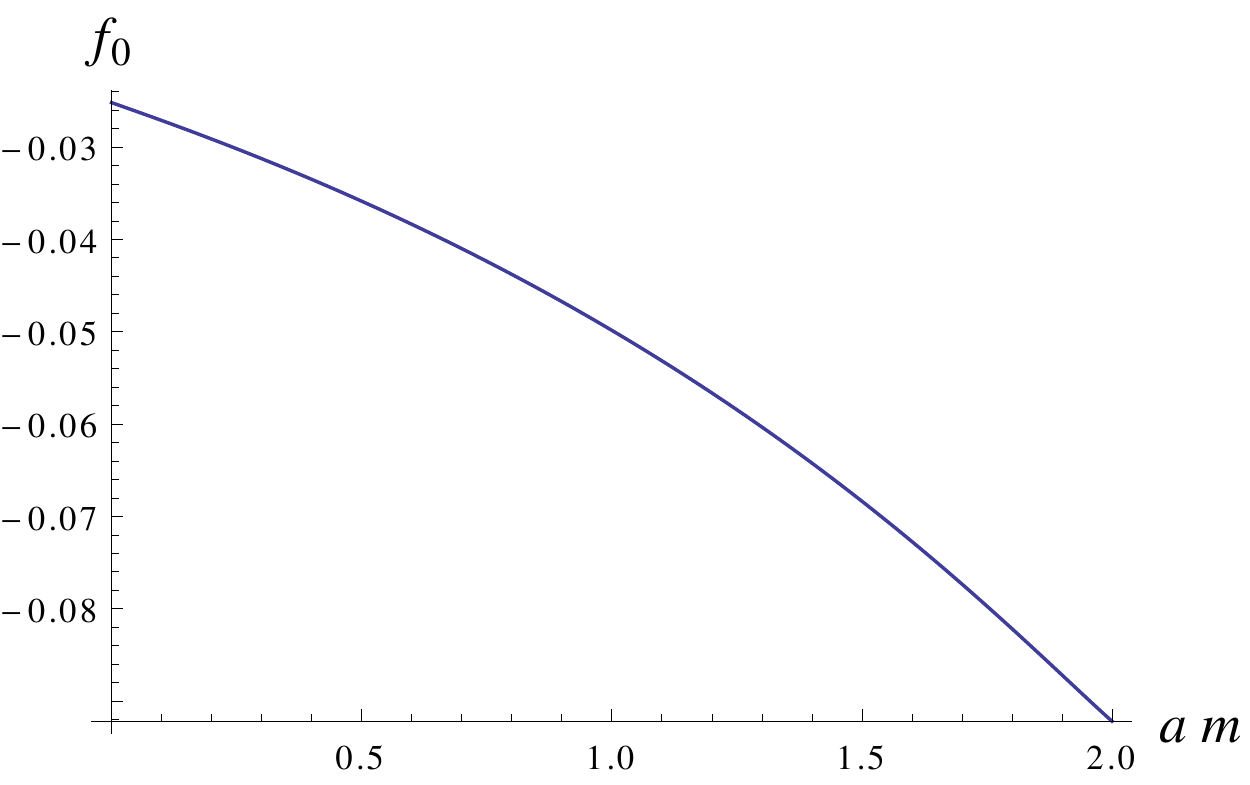}
  \caption{$f_0(am)$}
  \label{fig:1}
 \end{minipage}\hspace{3em}
 \begin{minipage}{0.45\columnwidth}
  \centering
  \includegraphics[width=\columnwidth]{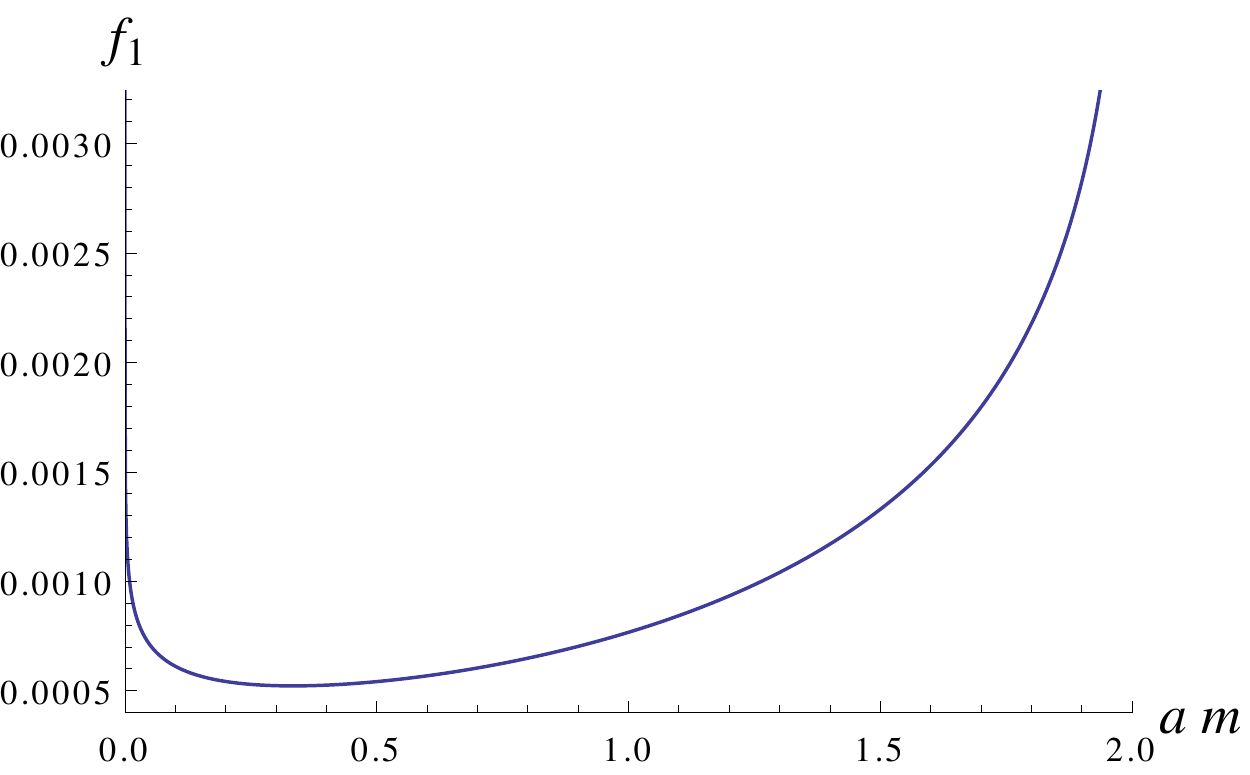}
  \caption{$f_1(am)$}
  \label{fig:2}
 \end{minipage}
\end{figure}

\begin{figure}[ht]
 \begin{minipage}{0.45\columnwidth}
  \centering
  \includegraphics[width=\columnwidth]{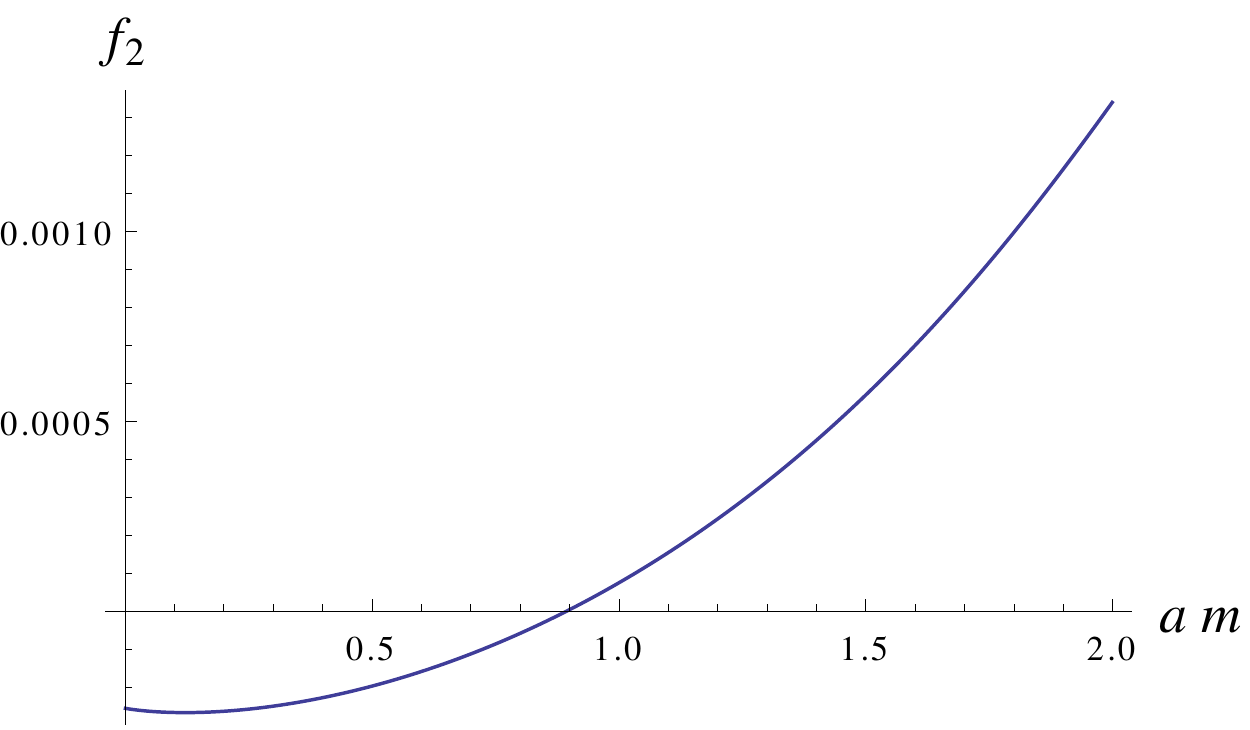}
  \caption{$f_2(am)$}
  \label{fig:3}
 \end{minipage}\hspace{3em}
 \begin{minipage}{0.45\columnwidth}
  \centering
  \includegraphics[width=\columnwidth]{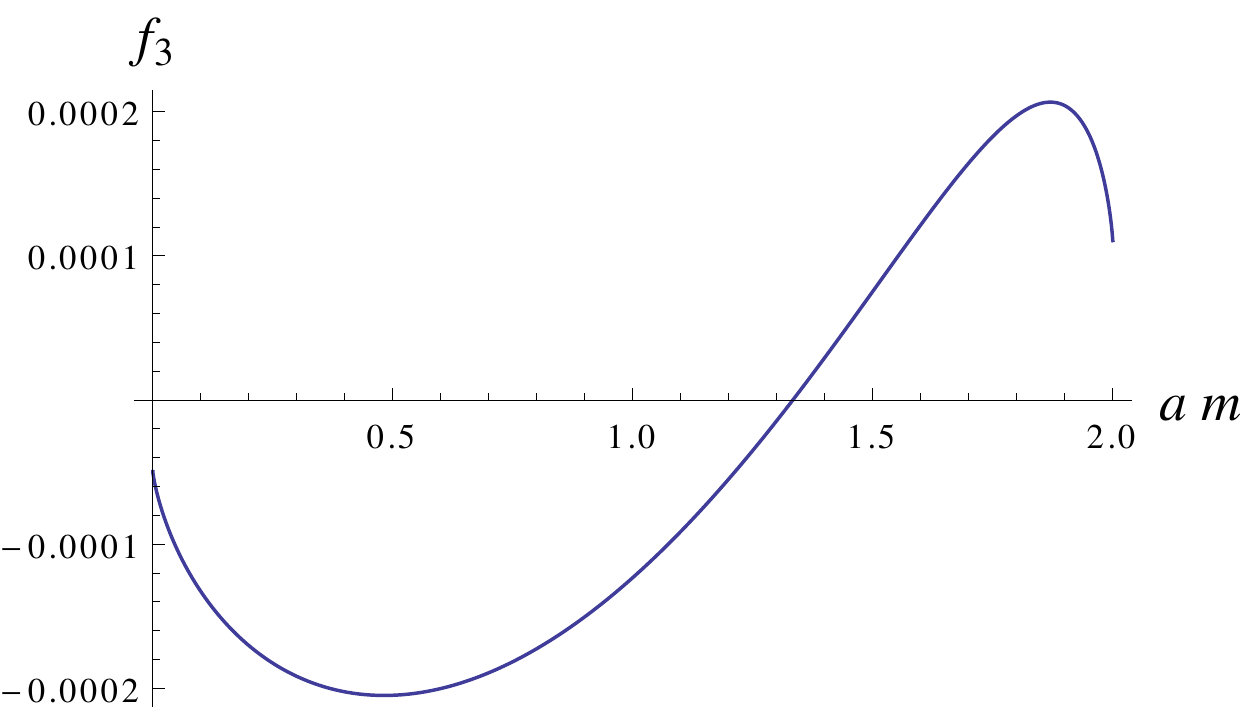}
  \caption{$f_3(am)$}
  \label{fig:4}
 \end{minipage}
\end{figure}

\begin{figure}[ht]
 \begin{minipage}{0.45\columnwidth}
  \centering
  \includegraphics[width=\columnwidth]{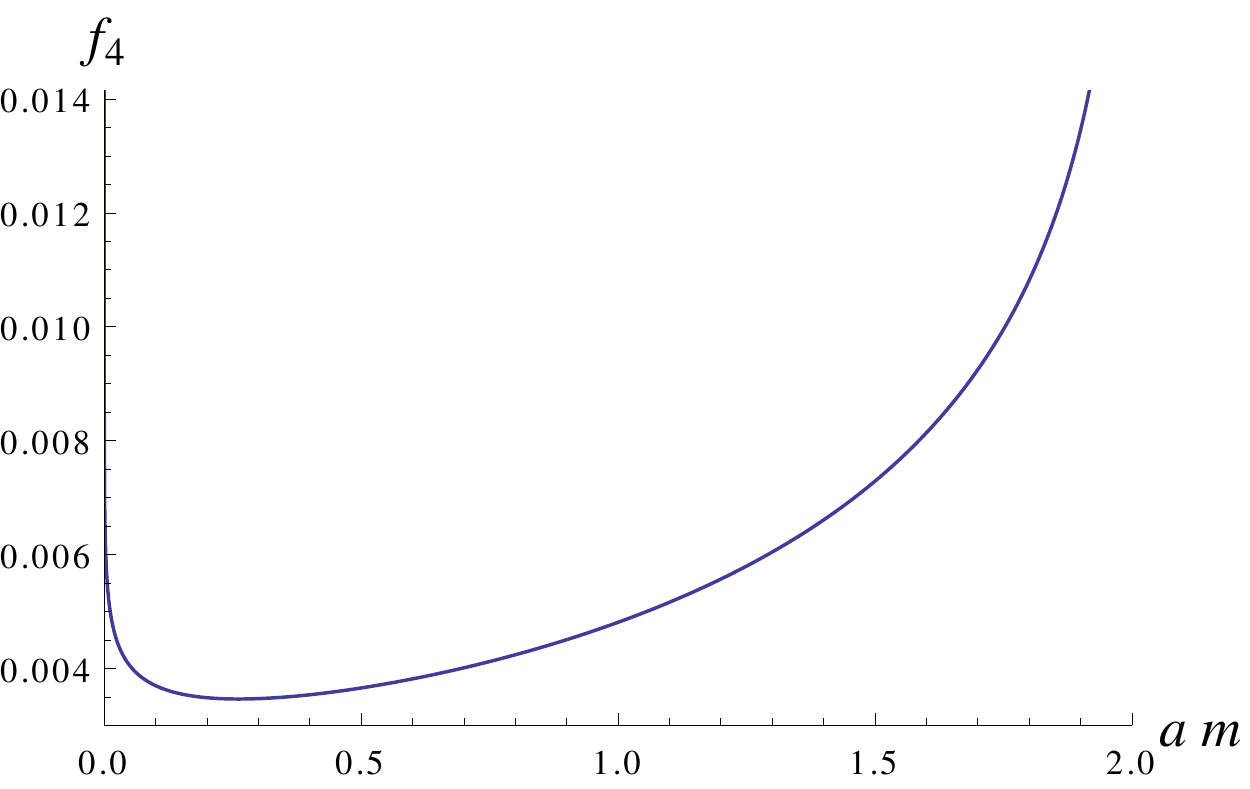}
  \caption{$f_4(am)$}
  \label{fig:5}
 \end{minipage}\hspace{3em}
 \begin{minipage}{0.45\columnwidth}
  \centering
  \includegraphics[width=\columnwidth]{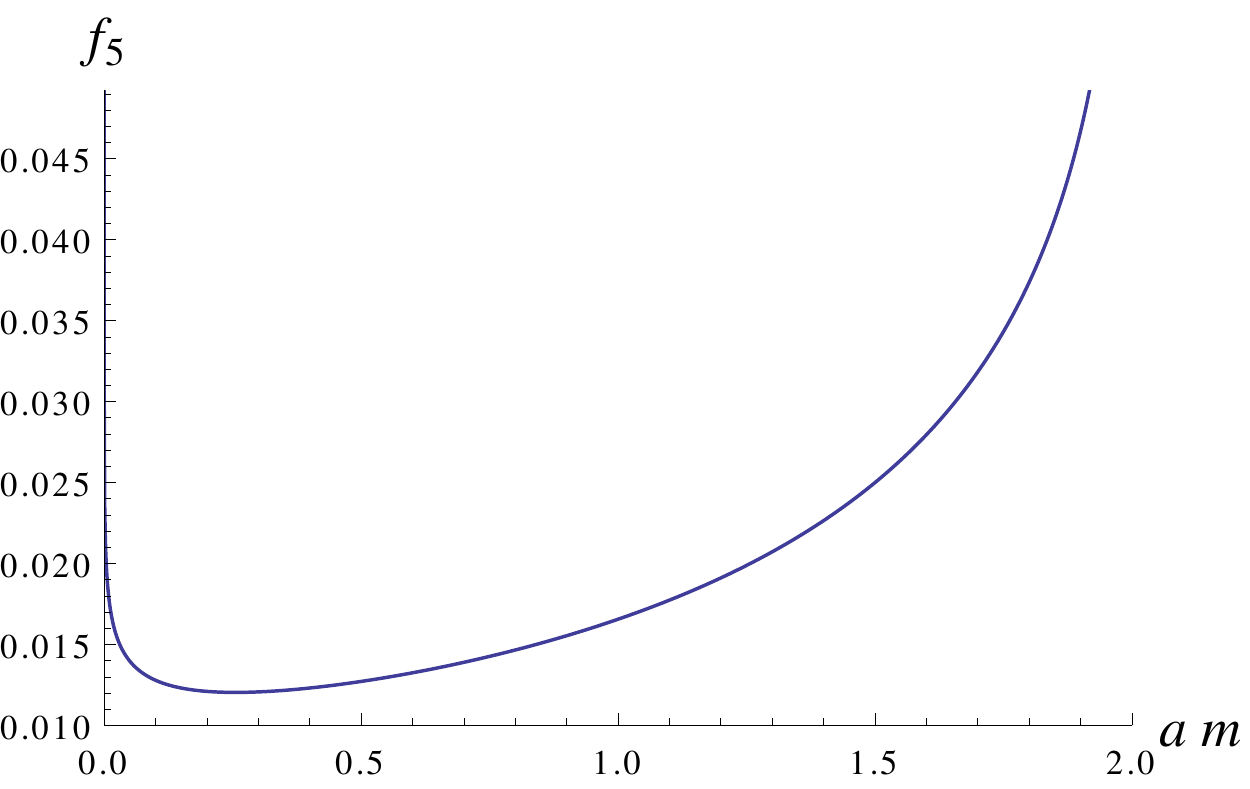}
  \caption{$f_5(am)$}
  \label{fig:6}
 \end{minipage}
\end{figure}

In this section, we show the results of the continuum limit
of~$\mathcal{L}(A, A_\star; \delta A, \delta_\star A_\star)$.\footnote{For
details of the computation of~Eq.~\eqref{eq:(1.8)},
see~Ref.~\cite{Makino:2016auf} and our work~\cite{Makino:2017pbq}.} In what
follows, we use the variables
\begin{align}
   C_\mu&\equiv A_{\star\mu}-A_\mu,
\label{eq:(2.1)}
\\
   \Bar{A}_\mu&\equiv\frac{1}{2}(A_\mu+A_{\star\mu}),
\label{eq:(2.2)}
\\
   \Bar{D}_\mu&\equiv\partial_\mu+[\Bar{A}_\mu,\cdot],
\label{eq:(2.3)}
\end{align}
and the field strength
\begin{align}
 \Bar{F}_{\mu\nu}&=\partial_\mu\Bar{A}_\nu-\partial_\nu\Bar{A}_\mu
 +[\Bar{A}_\mu,\Bar{A}_\nu] .
\label{eq:(2.4)}
\end{align}
We also define the following lattice integrals:
\begin{align}
 f_0(am)
 &\equiv
 \int_p\left(-\frac{1}{4t}-\frac{s_\rho^2}{4t}-\frac{c c_\rho}{4t}\right),
\label{eq:(2.5)}
\\
 f_1(am)
 &\equiv\int_p\left(\frac{1}{64t^2}-\frac{c_\rho c_\sigma}{128t}
 +\frac{s_\rho^2s_\sigma^2}{32t^2}\right),
\label{eq:(2.6)}
\\
 f_2(am)
 &\equiv
 \int_p\left(-\frac{c_\rho c_\sigma}{32t}+\frac{7s_\rho^2 s_\sigma^2}{64t^2}
 +\frac{cs_\rho^2c_\sigma}{32t^2}+\frac{c^2c_\rho c_\sigma}{64t^2}\right),
\label{eq:(2.7)}
\\
 f_3(am)
 &\equiv
 \int_p\left(-\frac{c_\rho c_\sigma}{32t}+\frac{3s_\rho^2s_\sigma^2}{32t^2}
 -\frac{s_\rho^2}{32t^2}-\frac{cc_\rho}{32t^2}\right),
\label{eq:(2.8)}
\\
 f_4(am)
 &\equiv
 \int_p\left(\frac{1}{96t}+\frac{s_\rho^2}{96t}
 +\frac{cc_\rho}{96t}+\frac{1}{16t^2}\right),
\label{eq:(2.9)}
\\
 f_5(am)
 &\equiv
 \int_p\left(\frac{1}{16t}+\frac{c_\rho c_\sigma}{32t}
 +\frac{7}{32t^2}-\frac{c^2}{32t^2}+\frac{c c_\rho}{16t^2}
 +\frac{s_\rho^2}{32t^2}\right),
\label{eq:(2.10)}
\end{align}
where
\begin{align}
   s_\rho&\equiv\sin p_\rho,&c_\rho&\equiv\cos p_\rho,
\label{eq:(2.11)}
\\
   c&\equiv\sum_\mu(c_\mu-1)+am,&t&\equiv\sum_\mu s_\mu^2+c^2,
\label{eq:(2.12)}
\\
   \int_p&\equiv\int_{-\pi}^\pi\frac{d^4p}{(2\pi)^4}.&&
\label{eq:(2.13)}
\end{align}
$f_i(am)$ ($i=0,\dots,5$) as the function of~$am$ are plotted
in~Figs.~\ref{fig:1}--\ref{fig:6}.

The local functional~$\mathcal{L}$ has three parts, according to the parity
and Lorentz symmetry: (i) the parity-odd and Lorentz-preserving part, (ii)
the parity-even and Lorentz-preserving part, and (iii) the parity-even and
Lorentz-violating part. First, the parity-odd part of~$\mathcal{L}$ is given
by
\begin{align}
   &\left.
   \mathcal{L}(A,A_\star;\delta A,\delta_\star A_\star)
   \right|_{\text{parity-odd}}
\notag\\
   &=-\frac{1}{32\pi^2}\epsilon_{\mu\nu\rho\sigma}
   \Biggl[\left(\Bar{F}_{\mu\nu}+\frac{1}{12}[C_\mu,C_\nu]\right)
   \{\delta A_\rho,\delta_\star A_{\star\sigma}\}
\notag\\
   &\qquad\qquad\qquad\qquad{}
   -\frac{1}{3}C_\mu
   \left(\{\delta A_\nu,\Bar{D}_\rho\delta_\star A_{\star\sigma}\}
   +\{\delta_\star A_{\star\nu},\Bar{D}_\rho\delta A_\sigma\}\right)
 \Biggr],
\label{eq:(2.14)}
\end{align}
where and in what follows the symbol~$\tr$ is assumed to be omitted. This
part is proportional to the gauge anomaly coefficient; thus this vanishes if
the anomaly cancellation condition is met. Second, we have the parity-even
and Lorentz-preserving part of~$\mathcal{L}$,
\begin{align}
   &\left.
   \mathcal{L}(A,A_\star;\delta A,\delta_\star A_\star)
   \right|_{\text{parity-even, Lorentz-preserving}}
\notag\\
   &=\frac{f_0}{a^2}\delta A_\mu\delta_\star A_{\star\mu}
\notag\\
   &\qquad{}
   +\left(-\frac{3f_1}{2}+\frac{f_2}{2}-\frac{f_3}{2}\right)
   [(\Bar{D}_\mu\delta A_\mu) C_\nu\delta_\star A_{\star\nu}
   -C_\mu\delta A_\mu(\Bar{D}_\nu\delta_\star A_{\star\nu})]
\notag\\
   &\qquad{}
   -\left(\frac{f_1}{2}+\frac{f_2}{2}-\frac{3f_3}{2}\right)
   [C_\mu(\Bar{D}_\nu\delta A_\mu)\delta_\star A_{\star\nu}
   -\delta A_\mu C_\nu(\Bar{D}_\mu\delta_\star A_{\star\nu})]
\notag\\
   &\qquad{}
   -\left(\frac{f_1}{2}+\frac{f_2}{2}-\frac{3f_3}{2}\right)
   [C_\nu\delta A_\mu(\Bar{D}_\mu\delta_\star A_{\star\nu})
   -(\Bar{D}_\nu\delta A_\mu)C_\mu\delta_\star A_{\star\nu}]
\notag\\
   &\qquad{}
   +\left(-\frac{7f_1}{2}+\frac{f_2}{2}+\frac{f_3}{2}\right)
   [(\Bar{D}_\mu C_\mu)\delta A_\nu\delta_\star A_{\star\nu}
   -\delta A_\nu(\Bar{D}_\mu C_\mu)\delta_\star A_{\star\nu}]
\notag\\
   &\qquad{}
   -\left(\frac{3f_1}{2}-\frac{f_2}{2}+\frac{f_3}{2}\right)
   [\delta A_\mu C_\mu(\Bar{D}_\nu\delta_\star A_{\star\nu})
   -C_\nu(\Bar{D}_\mu\delta A_\mu)\delta_\star A_{\star\nu}]
\notag\\
   &\qquad{}
   +\left(13f_1-3f_2-3f_3\right)
   (\Bar{D}_\mu\delta A_\mu)(\Bar{D}_\nu\delta_\star A_{\star\nu})
\notag\\
   &\qquad{}
   +\left(9f_1-3f_2-f_3\right)
   (\Bar{D}_\mu\delta A_\nu)(\Bar{D}_\mu\delta_\star A_{\star\nu})
\notag\\
   &\qquad{}
   +\left(-19f_1+5f_2+5f_3\right)
   (\Bar{D}_\nu\delta A_\mu)(\Bar{D}_\mu\delta_\star A_{\star\nu})
\notag\\
   &\qquad{}
   +\left(\frac{11f_1}{6}-\frac{f_2}{6}-\frac{7f_3}{6}\right)
   C_\mu\delta A_\nu C_\mu\delta_\star A_{\star\nu}
\notag\\
   &\qquad{}
   +\left(-\frac{13f_1}{6}+\frac{11f_2}{6}-\frac{7f_3}{6}\right)
   (C_\mu\delta A_\mu C_\nu\delta_\star A_{\star\nu}
   +C_\nu\delta A_\mu C_\mu\delta_\star A_{\star\nu})
\notag\\
   &\qquad{}
   +\left(-\frac{5f_1}{12}+\frac{19f_2}{12}-\frac{17f_3}{12}\right)
   (C_\nu C_\mu \delta A_\mu\delta_\star A_{\star\nu}
   +\delta A_\mu C_\mu C_\nu\delta_\star A_{\star\nu})
\notag\\
   &\qquad{}
   +\left(\frac{19f_1}{12}-\frac{5f_2}{12}-\frac{5f_3}{12}\right)
   (C_\mu C_\nu\delta A_\mu\delta_\star A_{\star\nu}
   +\delta A_\mu C_\nu C_\mu\delta_\star A_{\star\nu})
\notag\\
   &\qquad{}
   +\left(-\frac{17f_1}{12}+\frac{19f_2}{12}-\frac{11f_3}{12}\right)
   (C_\mu C_\mu\delta A_\nu\delta_\star A_{\star\nu}
   +\delta A_\nu C_\mu C_\mu\delta_\star A_{\star\nu}) ,
\label{eq:(2.15)}
\end{align}
and finally the parity-even and Lorentz-violating part is given by
\begin{align}
   &\left.
   \mathcal{L}(A,A_\star;\delta A,\delta_\star A_\star)
   \right|_{\text{parity-even, Lorentz-violating}}
\notag\\
   &=\frac{3}{2}\left(9f_1-f_2-f_3-\frac{f_4}{2}-\frac{f_5}{2}\right)
   [(\Bar{D}_\nu C_\nu)\delta A_\nu\delta_\star A_{\star\nu}
   -\delta A_\nu(\Bar{D}_\nu C_\nu)\delta_\star A_{\star\nu}]
\notag\\
   &\qquad{}
   -\left(9f_1-f_2-f_3-\frac{f_4}{2}-\frac{f_5}{2}\right)
   (\Bar{D}_\nu\delta A_\nu)(\Bar{D}_\nu\delta_\star A_{\star\nu})
\notag\\
   &\qquad{}
   +\left(\frac{47f_1}{2}-\frac{7f_2}{2}-\frac{7f_3}{2}
   +\frac{f_4}{4}-\frac{7f_5}{4}\right)
   C_\nu\delta A_\nu C_\nu\delta_\star A_{\star\nu}
\notag\\
   &\qquad{}
   +\left(\frac{67f_1}{4}-\frac{11f_2}{4}-\frac{11f_3}{4}
   +\frac{5f_4}{8}-\frac{11f_5}{8}\right)
   (C_\nu C_\nu\delta A_\nu\delta_\star A_{\star\nu}
   +\delta A_\nu C_\nu C_\nu\delta_\star A_{\star\nu}) .
\label{eq:(2.16)}
\end{align}

By using the above form
of~$\mathcal{L}(A,A_\star;\delta A,\delta_\star A_\star)$, one can deduce the
gauge variation of~$\ln\mathcal{Z}[A,0]$, $\delta^\omega\ln\mathcal{Z}[A,0]$
(see Appendix~A of~Ref.~\cite{Makino:2017pbq} for details). The parity-odd
part of~$\mathcal{L}$ gives rise to (leaving out the
symbol~$\int d^4 x\, \tr$)
\begin{equation}
   \left.\delta^\omega\ln\mathcal{Z}[A,0]\right|_{\text{parity-odd}}
   =\frac{1}{24\pi^2}\epsilon_{\mu\nu\rho\sigma}
   (\partial_\mu\omega)\left(A_\nu\partial_\rho A_\sigma
   +\frac{1}{2}A_\nu A_\rho A_\sigma\right),
\label{eq:(2.17)}
\end{equation}
which is the consistent gauge anomaly associated with a left-handed fermion.
It is impossible to rewrite this expression as the gauge variation of a local
term. On the other hand, the parity-even part
of~$\delta^\omega\ln\mathcal{Z}$ can be written as the gauge variation of
local terms:
\begin{align}
   &\left.\delta^\omega\ln\mathcal{Z}[A,0]
   \right|_{\text{parity-even}}
\notag\\
   &=\delta^\omega
   \Biggl[
   \frac{f_0}{2a^2}A_\mu A_\mu
\notag\\
   &\qquad\qquad{}
   +\frac{1}{2}
   (-13f_1+3f_2+3f_3)A_\mu\partial_\mu\partial_\nu A_\nu
\notag\\
   &\qquad\qquad\qquad{}
   +(5f_1-f_2-2f_3)
   (A_\mu\partial_\nu\partial_\nu A_\mu
   -A_\mu A_\nu\partial_\mu A_\nu
   +A_\mu A_\nu\partial_\nu A_\mu)
\notag\\
   &\qquad\qquad\qquad\qquad{}
   +\frac{2}{3}
   (f_1+f_2-2f_3)A_\mu A_\mu A_\nu A_\nu
   +\frac{1}{12}
   (-11f_1+f_2+7f_3)A_\mu A_\nu A_\mu A_\nu
\notag\\
   &\qquad\qquad\qquad\qquad\qquad{}
   +\frac{1}{2}
   \left(9f_1-f_2-f_3-\frac{f_4}{2}-\frac{f_5}{2}\right)
   A_\mu\partial_\mu\partial_\mu A_\mu
\notag\\
   &\qquad\qquad\qquad\qquad\qquad\qquad{}
   +\frac{1}{4}
   \left(19f_1-3f_2-3f_3+\frac{f_4}{2}-\frac{3f_5}{2}\right)
   A_\mu A_\mu A_\mu A_\mu
   \Biggr].
\label{eq:(2.18)}
\end{align}
The last two lines are not Lorentz invariant. This parity-even part does not
vanish even if the gauge representation is anomaly-free. For example, the
first term~$\delta^\omega[(f_0 / 2a^2) A_\mu A_\mu]$ corresponds to the gauge
variation of the mass term of the gauge field. The
\textit{regularization garbage} in~Eq.~\eqref{eq:(2.18)} can be subtracted by
local counterterms. However, such a necessity for counterterms will be
undesirable from a perspective of a non-perturbative formulation of chiral
gauge theories.

\begin{acknowledgement}
 We would like to thank Shoji Hashimoto, Yoshio Kikukawa,
 and Ken-ichi Okumura for
 valuable remarks.
 We are grateful to Ryuichiro Kitano and Katsumasa Nakayama for
 intensive discussions on a related subject.
\end{acknowledgement}

\appendix

\section{Gradient flow for infinite flow time}
\label{sec:A}

The gradient flow of the gauge field is defined by
\begin{equation}
 \partial_t B_\mu(t, x)
  = D_\nu G_{\nu\mu}(t, x), \qquad
  B_\mu(t=0, x) = A_\mu(x).
\label{eq:(A.1)}
\end{equation}
In the abelian theory,
we can solve this equation as
\begin{equation}
 B_\mu(t, x)
  = \int d^4 y \int \frac{d^4 p}{(2\pi)^4} e^{i p (x - y)}
  \left[\left(\delta_{\mu\nu} - \frac{p_\mu p_\nu}{p^2}\right)
   e^{- t p^2} + \frac{p_\mu p_\nu}{p^2} \right] A_\nu(y).
\label{eq:(A.2)}
\end{equation}
This shows that after infinite flow time the configuration becomes pure
gauge:
\begin{equation}
 B_\mu(t, x)
  \stackrel{t\to\infty}{\to} g(x)^{-1} \partial_\mu g(x),
\label{eq:(A.3)}
\end{equation}
where
\begin{equation}
 g(x)
  = \exp\left[- \int d^4 y \int \frac{d^4 p}{(2\pi)^4}
	 \frac{e^{i p (x - y)}}{p^2} \partial_\mu A_\mu(y)\right] .
\label{eq:(A.4)}
\end{equation}
Note that $g(x)$ is a non-local functional of the original gauge
field~$A_\mu(y)$.

For the non-abelian theory, we cannot solve the flow equation in a closed
form. However, we can show that the Euclidean action
integral~$S = \int d^4 x\,\frac{1}{4g_0^2} G_{\mu\nu}^a(x) G_{\mu\nu}^a(x)$
monotonically decreases along the flow. Since the minimum of the action
integral in the topologically trivial sector is given by a pure gauge
configuration, the flowed configuration in the topologically trivial sector
approaches a pure gauge configuration. In fact, the pure gauge configuration
\begin{equation}
 B_\mu(t, x)
  = g(x)^{-1} \partial_\mu g(x)
\label{eq:(A.5)}
\end{equation}
is a stationary solution of the flow equation,
$\partial_t B(t, x) = 0$.

\end{document}